\documentclass{PoS}

\def\bea{\begin{eqnarray}}
\def\eea{\end{eqnarray}}

\def\lmatrix{\left(\begin{array}}
	\def\rmatrix{\end{array}\right)}

\def\msbar{\overline{\rm MS\kern-0.5pt}\kern0.5pt}

\usepackage{cite}
\usepackage{verbatim}

\title{Status of a minimal composite Higgs theory}

\ShortTitle{Status of a minimal composite Higgs theory}

\author{Zoltan Fodor\\
        University of Wuppertal, Department of Physics, Wuppertal D-42097, Germany\\
        Juelich Supercomputing Center, Forschungszentrum Juelich, Juelich D-52425, Germany\\
        Eotvos University, Pazmany Peter setany 1, 1117 Budapest, Hungary 
        \email{fodor@bodri.elte.hu}}

\author{{Kieran Holland}\\
        University of the Pacific, 3601 Pacific Ave, Stockton CA 95211, USA\\
        Albert Einstein Center for Fundamental Physics, Bern University, Bern, Switzerland\\
        \email{kholland@pacific.edu}}

\author{\speaker{Julius Kuti}\\
        University of California, San Diego, 9500 Gilman Drive, La Jolla, CA 92093, USA\\
        \email{jkuti@ucsd.edu}}

\author{\speaker{Santanu Mondal}\\
        Eotvos University, Pazmany Peter setany 1, 1117 Budapest, Hungary\\
        MTA-ELTE Lendulet Lattice Gauge Theory Research Group, 1117 Budapest, Hungary\\
        \email{santanu@bodri.elte.hu}}

\author{Daniel Nogradi\\
        Eotvos University, Pazmany Peter setany 1, 1117 Budapest, Hungary\\
        MTA-ELTE Lendulet Lattice Gauge Theory Research Group, 1117 Budapest, Hungary\\
        \email{nogradi@bodri.elte.hu}}

\author{\speaker{Chik Him Wong}\\
        University of Wuppertal, Department of Physics, Wuppertal D-42097, Germany\\
        \email{cwong@uni-wuppertal.de}}

\abstract{We analyze three sets of gauge ensembles in our extended physics program of a  particularly
	important BSM gauge theory with a fermion  doublet in the two-index symmetric (sextet) representation 
	of the SU(3) BSM color gauge group. Our investigations include chiral
symmetry breaking $\rm{(\chi SB)}$ in the p-regime and $\epsilon$-regime, the mass of 
the composite ${\rm 0^{++}}$ scalar, resonance spectroscopy, 
new physics from gauge anomaly constraints,
and the role of stable sextet BSM baryons with 
Electroweak interactions 
in dark matter searches.
Important new goals include studies of the 
${\rm 0^{++}}$  scalar entangled with Goldstone dynamics in the p-regime and the $\epsilon$-regime,
the resonance spectrum with particular attention to emerging LHC signals, like recent hints for 
diphoton excess at 750 GeV
 or diboson anomalies in the 2 TeV range. 
	All results reported here are 
	preliminary before journal publication including some post-conference material for the discussion.
 }

\FullConference{The 33rd International Symposium on Lattice Field Theory,
		 14 -18 July  2015\\
		 Kobe International Conference Center, Kobe, Japan}

\begin{document}
	
\section{Introduction}


An important strongly coupled near-conformal gauge theory built on the minimally required
SU(2) bsm-flavor doublet of two massless fermions, with a
confining gauge force at the TeV scale in the sextet
representation of the new SU(3) BSM color gauge group is an intriguing possibility 
for the minimal realization of the composite Higgs mechanism.
Early discussions of the model as a BSM candidate were initiated in
systematic explorations of higher fermion representations of color gauge groups~\cite{Dietrich:2005jn,Sannino:2004qp,Hong:2004td}
for extensions of the original Higgsless Technicolor paradigm~\cite{Susskind:1978ms,Weinberg:1979bn}.
In fact, the first appearance of the particular two-index symmetric SU(3) fermion representation 
can be traced even further back to Quantum Chromodynamics (QCD) where a doublet of sextet quarks
was proposed as a mechanism for Electroweak symmetry breaking (EWSB)
without an elementary Higgs field~\cite{Marciano:1980zf}. This idea had to be
replaced by a new
gauge force at the TeV scale, orders of magnitude stronger than in QCD, to facilitate the dynamics 
of EWSB just below the lower edge of the conformal window in 
the new BSM paradigm~\cite{Dietrich:2005jn,Sannino:2004qp,Hong:2004td}.
It should be noted that throughout its early history the important near-conformal behavior 
of the model was not known and definitive results had to wait for recent
non-perturbative investigations with lattice gauge theory methods
as used in our work.

Near-conformal BSM theories raise
the possibility of a light composite scalar, perhaps a Higgs impostor, to emerge from new strong dynamics,
far separated from the associated composite resonance spectrum in the few TeV mass range
with interesting and testable predictions for the Large Hadron Collider (LHC). 
This scenario is very different from what was expected from QCD when scaled up to the Electroweak scale,
as illustrated by the failure of the Higgsless Technicolor paradigm.
Given the discovery of the 125 GeV Higgs particle at the LHC, any realistic BSM theory must contain a Higgs-like state, 
perhaps with some hidden composite structure.

Based on our {\it ab initio} non-perturbative  lattice calculations
we find
accumulating evidence for near-conformal behavior in the sextet theory with
the emergent low mass  $0^{++}$ scalar state far separated from 
the composite resonance spectrum of bosonic and baryonic excitations in the 2-3 TeV energy range~\cite{Fodor:2012ty,Fodor:2012ni,Fodor:2014pqa,Fodor:2015vwa}. 
The identification of the light scalar state is numerically challenging since it
requires the evaluation of disconnected fermion loop contributions to correlators with vacuum quantum numbers
in the range of light fermion masses we explore.
The evidence to date is very promising that the $0^{++}$ scalar is light 
in the chiral limit and that the model at this stage remains an important BSM candidate.
This report presents new results and outlines the need for important further work needed for definitive results.

In Section 2 for the first time we describe the anomaly-free 
Electroweak embedding of the strongly coupled sextet gauge theory in a more comprehensive way 
than before and present the need for more extensive baryon analysis of our gauge ensembles with 
relevance for dark matter searches and model viability.
In Section 3 new results are reported and analysis tools are proposed for mass-deformed chiral 
perturbation theory and its chiral limit using p-regime and $\epsilon$-regime technologies including
the improved chiral properties of mixed actions in the valence sector.
This is critically important for accurate BSM scale setting 
from the chiral limit of the Goldstone decay constant ${\rm F_\pi}$ and the chiral condensate .
In Section 4 the status of the light ${\rm 0^{++}}$ scalar and the resonance spectrum is discussed
with new plans briefly outlined. Section 5 describes efforts to understand physics at several scales in 
nearly conformal gauge theories from the asymptotically free UV regime with an almost conformal crossover 
to the infrared with spontaneous chiral symmetry breaking. Flavor dependence of the running coupling and beta function 
in the sextet and fundamental fermion representations illustrates this problem. 
In Section 6 we provide a brief summary of the computational aspects 
of our work describing the staggered fermion action
we use in the simulations with well known taste breaking effects on the exact chiral symmetries of the continuum theory.
All figures are attached at the end of the report without interrupting the flow of the narrative.

\section{Electroweak multiplet structure, gauge anomalies, and baryons}

As in the minimal scheme of Susskind~\cite{Susskind:1978ms} and Weinberg~\cite{Weinberg:1979bn}, 
the gauge group of the theory  is 
${\rm SU(3)_{bsm}\!\otimes\!SU(3)_c\!\otimes\!SU(2)_w\!\otimes\!U(1)_Y}$ where  ${\rm SU(3)_c}$ designates 
the QCD color gauge group and ${\rm SU(3)_{bsm}}$ represents the BSM color gauge group 
of the new strong gauge force.
In addition to quarks and leptons of the Standard Model, we include one ${\rm SU(2)}$ 
bsm-flavor doublet ${\rm  (u,d)}$ of fermions which are ${\rm SU(3)_c}$ singlets and transform 
in the six-dimensional sextet representation of BSM color, distinct from
the fundamental color representation of fermions in the original Technicolor
scheme~\cite{Susskind:1978ms,Weinberg:1979bn}. 
The formal designation ${\rm  (u,d)}$ for the bsm-flavor doublet of sextet fermions uses a similar notation 
to the two light quarks of QCD but describes completely different physics. 
The massless sextet fermions form two chiral doublets 
${\rm (u,d)_L}$ and ${\rm (u,d)_R}$ under the global symmetry group
${\rm SU(2)_L\!\otimes\!SU(2)_R\otimes\!U(1)_B}$. Baryon number
is conserved 
for quarks of the Standard Model separate from baryon number conservation 
for sextet fermions which carry $1/3$ of BSM baryon charge associated 
with the BSM sector of the global ${\rm U(1)_B}$  symmetry group.
%

\subsection{Electroweak multiplet structure}

It is straightforward to define 
consistent multiplets  for the sextet fermion flavor doublet 
under the ${\rm SU(2)_w\!\otimes\!U(1)_Y}$ Electroweak gauge group
with hypercharge assignments for left- and right-handed fermions 
transforming under the ${\rm SU(2)_w}$  weak isospin group.
The two fermion flavors ${\rm {u^{ab}}}$ and ${\rm d^{ab}}$ of the strongly coupled sector
carry six colors in two-index symmetric tensor notation,
${\rm a,b = 1,2,3}$,  associated with the gauge force of the ${\rm SU(3)_{bsm}}$ group.
This is equivalent to a six-dimensional vector notation in the sextet representation.
The fermions transform as left-handed weak isospin doublets
and right-handed weak isospin singlets for each color,
\begin{equation}
{\rm \psi^{ab}_L = \lmatrix{c}  {\rm u^{ab}_L} \\  {\rm d^{ab}_L }\rmatrix , \qquad \psi^{ab}_R = (u^{ab}_R ,\; d^{ab}_R) }.
\end{equation}
With this choice of representations, the normalization for the hypercharge ${\rm Y}$ 
of the ${\rm U(1)_Y}$ gauge group is 
defined by the relation ${\rm Y=2(Q-T_3)}$, with  ${\rm T_3}$ designating the third component of weak isospin.

Once Electroweak gauge interactions are turned on, 
the chiral symmetry breaking pattern  ${\rm SU(2)_L\!\otimes\!SU(2)_R \rightarrow SU(2)_V}$ 
of strong dynamics breaks  Electroweak symmetry in the expected pattern,
${\rm SU(2)_w \times U(1)_Y \rightarrow U(1)_{\rm em}}$, and with the simultaneous
dynamical realization of the composite Higgs mechanism. It is important to note that the 
dynamical Higgs mechanism is facilitated through the electroweak gauge couplings of
the sextet fermions and does not depend on the hypercharge assignments 
of the multiplets~\cite{Susskind:1978ms}. 
Recently, we presented a detailed analysis on anomaly constraints of hypercharge assignments~\cite{Fodor:2016wal}.
In this report we summarize what is relevant for the sextet baryon analysis of our existing gauge ensembles 
with dark matter and model viability implications~\cite{Fodor:2016wal}.

\subsection{Anomaly conditions}

Anomaly constraints have a long history in Technicolor motivated BSM model building with 
representative examples 
in~\cite{Dietrich:2005jn,Kainulainen:2006wq,Foadi:2007ue,Antola:2009wq,Kainulainen:2009rb}.
The first condition for model construction with left-handed doublets is the global Witten anomaly 
constraint which requires an even number of left-handed SU(2) multiplets to 
avoid inconsistency in the theory from a vanishing fermion determinant of the partition function~\cite{Witten:1982fp}.
In addition, gauge anomaly constraints also have to be satisfied~\cite{Adler:1969er}. 
With vector current ${\rm  V^i_\mu(x)=\overline{\psi}T^i\gamma_\mu\psi(x)}$ and axial current 
${\rm A^i_\mu(x)=\overline{\psi}T^i\gamma_\mu\gamma_5\psi(x)}$
constructed from fermion fields and internal symmetry matrices ${\rm T^i}$ in some group representation 
R for fermions, the anomaly in the axial vector Ward
identity is proportional to ${\rm tr(\{T^i(R),T^j(R)\}T^k(R))}$ and must vanish. 
In the sextet theory fermions are either left-handed doublets or right-handed singlets 
under the ${\rm SU(2)_w}$ gauge group. 
The matrices ${\rm T^i}$  will be either the ${\rm \tau^i}$ Pauli matrices 
or the diagonal ${\rm U(1)}$ hypercharge ${\rm Y}$.
Since the ${\rm SU(2)}$  group is  anomaly free, ${\rm tr(\{\tau^i,\tau^j\}\tau^k)=0}$, 
we only need to consider anomalies where at least one ${\rm T^i}$ is the hypercharge Y.
The non-trivial constraints come from two conditions on hypercharge traces,
\begin{equation}
{\rm tr(Y )=0, \quad tr(Y^3) \propto tr(Q^2T_3 - QT^2_3) = 0}\;, 
\label{eq:hyper}
\end{equation}
where  ${\rm Y=2(Q-T_3)}$ with electric charge ${\rm Q}$, and 
${\rm T_3}$ as the third component of weak isospin. 
There are two simple solutions for BSM model building with sextet fermions to satisfy 
the Witten anomaly condition and
gauge anomaly constraints on tr(Y) and ${\rm tr(Y^3)}$ in Eq.~(\ref{eq:hyper}). 
The first solution with the choice ${\rm Y(f_L)=0}$ for doublets of left-handed sextet fermions ${(\rm f_L)}$
leads  to half-integer electric charges for composite baryons. The second solution
with the  choice ${\rm Y(f_L)=1/3}$ for doublets of left-handed sextet fermions leads to integer electric charges
for composite baryons.  The hypercharges of right-handed singlets are automatically set from
consistent electric charge assignments in both cases. The two choices have very different 
implications for sextet baryons.

\subsection{Sextet baryons and their Early Universe}

In the sextet BSM theory we do not have direct observations of new heavy baryons
to set unique  hypercharge
assignments for left-handed doublets and right-handed singlets of sextet fermions from two alternate 
solutions to the anomaly conditions. 
Viability of the choices  ${\rm Y(f_L)=0}$, or  ${\rm Y(f_L)=1/3}$,  is
affected by the different electric charge assignments they imply. 
With heavy sextet baryon masses in the 3 TeV range, as determined from our recent lattice 
simulations~\cite{Fodor:2016wal},
the seemingly minimal solution with ${\rm Y=0}$ for left-handed doublets would lead to intriguing 
predictions of baryon states with half-integer electric charges for future accelerator searches and relics
with fractional electric charges from the early Universe with observable consequences.
Problems with half-integer electric charges, from the choice ${\rm Y(f_L)=0}$ in our case, were anticipated earlier 
from strong observational limits on stable fractional charges in the early Universe 
and their terrestrial relics~\cite{Chivukula:1989qb,Langacker:2011db}.
The non-controversial ${\rm Y(f_L)=1/3}$ anomaly solution for the sextet model has 
new dark matter implications~\cite{Fodor:2016wal} which require new lattice calculations proposed here.

\subsection{ Ongoing lattice work on sextet baryons and future plans}

The lightest baryons in the strongly coupled sextet gauge sector are expected to form 
isospin flavor doublets  ${\rm (uud,udd)}$, similar to the pattern in QCD.
As we noted earlier, baryons in the sextet model  
should carry integer multiples of electric charges if ${\rm Y(f_L)\neq 0}$ to avoid problems with the relics
of the early Universe. This leads to the
simplest choice  ${\rm Y(f_L)=1/3}$ with gauge anomalies to be compensated. 
A new pair of left-handed lepton 
doublets emerged from this choice as the simplest manifestation 
of the anomalies and the Electroweak extension of the strongly coupled sextet gauge sector~\cite{Fodor:2016wal}.

Neutron-like ${\rm udd}$ sextet model baryons ${\rm (n_6)}$ will carry no electric charge 
and proton-like ${\rm uud}$ sextet model baryons ${\rm (p_6)}$ have one unit of positive electric charge 
from the choice ${\rm Y(f_L)=1/3}$. 
The two baryon masses are split by electromagnetic interactions.
The ordering of the two baryon masses in the chiral limit of massless sextet fermions will
require non-perturbative {\em ab initio} lattice calculations of the electromagnetic 
mass shifts to confirm intuitive expectations
that the neutron-like ${\rm n_6}$ baryon has lower mass than the proton-like ${\rm p_6}$ baryon.
In QCD this pattern was confirmed by recent lattice calculations~\cite{Borsanyi:2014jba}.
We expect the same ordering in the sextet model so that the proton-like ${\rm p_6}$  baryon 
will decay very fast, ${\rm p_6 \rightarrow n_6 + ... }$,  with a lifetime ${\rm \tau \ll 1~second}$. 
It is unlikely for rapidly decaying  ${\rm p_6}$ baryons to leave any relic footprints 
from dark nucleosynthesis before they decay.

With BSM baryon number conservation the neutral ${\rm n_6}$ baryon is stable 
and observational limits on its direct detection from experiments like 
XENON100~\cite{Aprile:2012nq} and LUX2013~\cite{Akerib:2013tjd} have to be estimated.
In charge symmetric thermal evolution sextet model baryons are produced with relic number density ratio
${\rm  n_{B_6}/n_B  \approx 3\cdot 10^{-7}}$.
For 3 TeV sextet model baryon masses we can estimate the detectable dark matter ratio of 
respective mass densities ${\rm \rho_{B_6}}$ and ${\rm \rho_{B}}$ as
${\rm \rho_{B_6}/\rho_B  \approx 10^{-4}}$,
about ${\rm 5\cdot 10^{4}}$ times less than the full amount of unaccounted dark mass, 
${\rm \rho_{dark} \approx 5\cdot \rho_B}$.  
We will use this mass density estimate to guide observational limits on relic sextet model baryons
emerging from charge symmetric thermal evolution where
tests of dark baryon detection come from elastic collisions with nuclei in dark matter detectors~\cite{Fodor:2016wal}.
The neutral and stable ${\rm n_6}$ 
baryon can interact several different ways with heavy nuclei in direct detection experiments
including (a) magnetic dipole interaction, (b) Z-boson exchange, (c) Higgs boson exchange, and 
(d) electric polarizability.
It turns out that cross sections from (a) and (b) can be parametrized and estimated even without 
lattice simulations. Cross sections from (c) and (d) require lattice calculations using our existing gauge ensembles
and capacity computing from new allocation we request for gpu capacity computing.
Based on these estimates we expect to show that the sextet BSM model is consistent with observational limits and 
stable baryons
will contribute a small fraction to the missing dark matter content. 
New physics implied by gauge anomaly constraints, like new lepton generations with neutrinos~\cite{Fodor:2016wal}, 
can also contribute to the relic abundance of dark matter.
These are  important and interesting issue for future investigations.

\section{Mass-deformed chiral perturbation theory and the chiral condensate}
One of the most important goals of lattice BSM models is to accurately set the Electroweak scale 
as a function of the lattice spacing. 
This allows control on the continuum limit when the cutoff is removed and phenomenologically relevant 
BSM predictions are made.
The chiral ${\rm SU(2)_L\times SU(2)_R}$ symmetry of the model is dynamically broken to the diagonal vector 
symmetry ${\rm SU(2)_V}$ and 
three associated Goldstone pions facilitate
the minimal realization of the Higgs mechanism after the Electroweak interactions are turned on.
The Electroweak scale in finite 
lattice spacing units is set from the decay constant ${\rm F_\pi}$ of the Goldstone pion in the 
chiral limit with ${\rm F=250~GeV}$ in continuum physics notation. 
It can be identified as the fundamental scale of the theory related to the chiral (Higgs) condensate 
through the GMOR relation.

\subsection{Taste breaking cutoff effects in the staggered pion spectrum}

Since the determination of the Goldstone decay constant ${\rm F}$ in the chiral limit is critically important 
for the location of the light scalar mass and the well-separated resonance spectrum in the 2-3 TeV range, 
we carefully monitor taste breaking effects in the pion spectrum with the goal of removing cutoff effects from physics predictions. This also  serves as guidance  
for our choice of lattice spacings for new configuration generation. 

To illustrate cutoff dependent taste breaking effects, spectra of mass-deformed non-Goldstone pion states are shown
in Figure~\ref{fig:goldstone} from our newest data with the definition of the relevant correlators and
quantum numbers given in~\cite{Fodor:2011tu,Fodor:2012ty}. In the fermion mass range of our data set the taste breaking pattern 
is different from QCD where the residual ${\rm \Delta}$ mass shifts of the non-Goldstome pions are equispaced
in the chiral limit with approximately degenerate SO(4) taste multiplets and with  parallel slopes for finite fermion mass deformations 
of Goldstone and non-Goldstone pion states~\cite{Lee:1999zxa}.
For example, as part of the equispaced split of degenerate SO(4) multiplets, the observed approximate split
${\rm \Delta_{ij} \sim  2\Delta_{sc}}$ of two multiplets in QCD appears to have collapsed in the sextet model.
The other distinct difference from QCD is the non-parallel slopes which fan out  
in Goldstone and non-Goldstone  mass deformations of the pion spectrum as shown
in Figure~\ref{fig:goldstone}. While the ${\rm \Delta}$ additive mass shifts are LO taste breaking effects 
in the chiral Lagrangian~\cite{Lee:1999zxa,Aubin:2003mg}, the taste breaking slope corrections ${\rm \delta}$
can plausibly be identified with NLO analytic terms in rooted staggered chiral perturbation 
theory (${\rm rs\chi PT}$)~\cite{Sharpe:2004is}. The corrected mass 
relation is ${\rm M^2_{NLO} = M^2_{LO}(1+\delta)}$
where ${\rm \delta}$ depends on the taste quantum number of the pion state. 
Several relations
constrain the ${\rm \delta}$ taste breaking corrections~\cite{Sharpe:2004is}.
The pion spectrum with taste breaking cutoff effects is the input to analyze the fundamental 
parameters of ${\rm rs\chi PT}$  as worked out for the SU(3) group 
in~\cite{Aubin:2003mg}.
Our adaptation to the SU(2) group of ${\rm rs\chi PT}$  in the sextet model is straightforward.

\subsection{Fundamental parameters from rooted staggered chiral perturbation theory (p-regime)}

For the SU(2) analysis we adapted the procedure from~\cite{Aubin:2003mg}.
There are two fundamental parameters F and B in the SU(2) chiral Lagrangian. The fundamental parameter F
of ${\rm \chi PT}$, defined as the chiral limit of
the pion decay constant ${\rm F_\pi}$, sets the Electroweak scale and the fundamental parameter  B sets the 
fermion mass deformation of the Goldstone spectrum. With bare fermion mass m, the 
RG invariant combination ${\rm m\cdot B F^2}$ is related 
to the chiral condensate via the GMOR relation. 

We apply rooted staggered chiral perturbation theory to the mass-deformed pion spectrum and ${\rm F_\pi}$. 
The fitting procedure in the p-regime proceeds in several steps. In the first step  
finite volume correction is applied to the ${\rm M_\pi}$ and ${\rm F_\pi}$ data from 1-loop continuum ${\rm \chi PT}$.
This is sufficient to assure that in the next step the fitting procedure 
is applied to data free from volume dependence. 
A linear fit is applied to the quadratic masses of the non-Goldstone pion spectrum to determine their 
mass shifts and slopes.
In the final analysis of rooted chiral perturbation theory, non-Goldstone pion states run in the chiral loops 
including their mass splittings and fan-out slope structure from taste breaking as determined from the  
linear fits to the non-Goldstone spectrum. 
We applied this analysis at two values of the gauge coupling where we have extensive ensembles. 

For illustration, preliminary results from ${\rm rs\chi PT}$ are shown in Figure~\ref{fig:chipt}
from fits at gauge coupling ${\rm \beta = 3.20}$ 
which corresponds to our coarser lattice of the two extended sets of gauge ensembles. 
The upper left panel shows the linear fits to the quadratic masses of the non-Goldstone pions to determine their 
mass shifts and slopes as input. The upper right panel shows the 
${\rm rs\chi PT}$ fit to ${\rm F_\pi}$ as a function of fermion mass
deformations away from the chiral limit. The two lower panels  show  
${\rm rs\chi PT}$ fits to ${\rm M_\pi}$ as a function of fermion mass
deformations away from the chiral limit. 
Fits at the finer lattice 
spacing ${\rm \beta = 3.25}$ are quite similar in quality but with lower confidence level. 
The unambiguous determination of the cutoff dependent F and B parameters and their continuum limit 
from  ${\rm rs\chi PT}$ will require extended analysis.
Partial quenching with valence fermions is the first added step to make the ongoing analysis more robust.

Although our results are consistent with chiral symmetry breaking and ${\rm rs\chi PT}$,  ongoing work will 
require considerable  extensions for definitive results. Important new work, besides partial quenching 
includes a solution to the entanglement problem of the
light scalar with the low pion spectrum in perturbation theory with comparable masses in the 
${\rm rs\chi PT}$  regime, crossover analysis from the p-regime to the $\epsilon$-regime and 
applications of Random Matrix Theory (RMT) in the  $\epsilon$-regime. Coupled chiral dynamics of the low 
mass scalar ${\rm 0^{++}}$ state with the pions requires new analysis based on an extended effective theory. 
We are using the modified effective field theory of ${\rm \chi PT}$ on existing gauge ensembles
but new gauge configuration generation is also needed in the crossover to the $\epsilon$ regime.
For independent control on the results for the fundamental parameters F and B we developed and apply now 
mixed actions with improved chiral symmetry without taste breaking in the valence sector of the analysis. 
Ongoing new efforts in the p-regime and RMT based $\epsilon$-regime  in mixed action setting
will resolve important aspects of ${\rm rs\chi PT}$ with better determination of F and B.

\subsection{Epsilon-regime,  RMT, and mixed action in the valence sector}

Safe extrapolation from the entangled regime of the low mass ${\rm 0^{++}}$  scalar with pions to the massless fermion limit
is enabled by crossover to the $\epsilon$-regime of ${\rm \chi PT}$ at low enough scales ${\rm \lambda}$ where Goldstone dynamics begins to decouple from the scalar state.  This is demanding and requires significant resources. 
To control taste breaking we cannot go to lattice spacings coarser than the one set by ${\rm \beta=3.20}$. 
The uncertainties in the value of ${\rm F\sim 0.018-0.025}$ with limitations from rooted chiral perturbation theory at this 
lattice spacing requires large ${\rm V=56^3\times 96}$ and ${\rm V=48^3\times 96}$  lattice volumes 
to control the ${\rm F\cdot L \geq 1}$ 
condition which is necessary for convergent expansion in all regimes of ${\rm \chi PT}$, including the $\epsilon$-regime. 
Even for our largest ${\rm V=56^3\times 96}$ and ${\rm V=48^3\times 96}$  lattice volumes control with
${\rm F\cdot L \sim 1}$ is just barely sufficient. 
For the lowest fermion mass  ${\rm m=0.0010}$,  we have now at these volumes and at this lattice spacing, 
the scaling variable ${\rm m\Sigma V\sim 80}$ is very large and more appropriate for the p-regime analysis
of ${\rm \chi PT}$. Reaching the $\epsilon$-regime requires substantial decrease in 
the scaling variable ${\rm m\Sigma V}$ targeting ${\rm m=0.0003}$
which presents considerable algorithmic challenge for accelerated inversion methods and also calls for 
mixed action innovation. 
We deploy accelerated inverters in configuration generation to the  ${\rm m=0.0010-0.0003}$ 
range and analyze these configurations with mixed valence actions of good chiral properties as described below. 
Our limited resources this year allowed us to test these methods without comprehensive deployment for 
phenomenologically relevant results like the ${\rm M_{0^{++}}/F}$ ratio in the continuum limit.

The ${\rm m=0.0010-0.0003}$ range is in the crossover from the p-regime to the $\epsilon$-regime
where known methods of ${\rm \chi PT}$ are based on partial quenching and mixed action analysis.
For reliable testing, we performed ${\rm \chi PT}$ analysis 
in the crossover to the $\epsilon$-regime with partial quenching and a  mixed valence action with 
improved chiral symmetries. 
We take the  p-regime gauge configurations of the lowest fermion masses on the largest lattice volumes
and analyze the fermion condensate and the Dirac spectrum  with valence fermion action 
where the original gauge link variables are replaced with the ones with a fixed number of small stout steps 
which corresponds to fixed gradient flow time ${\rm t}$ in lattice spacing units at each gauge coupling. 
This strategy 
can be viewed as a mixed action based analysis with very good chiral properties of the fermion valence action.

The newest test results are shown in Figure~\ref{fig:mixed}.
The valence action is defined with a large number of very small stout steps  
which corresponds to  gradient flow time ${\rm t=3}$ in
cutoff units. We checked the eigenvalues of the Dirac operator which order into nearly degenerate quartets 
with the smeared gauge links of the gradient flow. The degenerate eigenvalues follow the index theorem count matching
the topology of each gauge configuration as measured from the topological charge operator on the gradient flow. 
The upper left panel shows the RMT distribution of four degenerate quartets showing that
the $\epsilon$-regime is reached with the scaling variable ${\rm \lambda\Sigma_{mixed} V\sim 10}$ 
where the fermion mass is replaced by the scale 
of the gradient flow defined valence Dirac spectrum (${\rm m\rightarrow\lambda)}$.
The fermion condensate  ${\rm \Sigma_{mixed}}$, not RG invariant itself, is 
consistently determined from the gradient flow defined valence Dirac operator.
The upper right panel illustrates the perfect degeneracy of the Goldstone pion with one selected non-Goldstone pion 
(scPion in the plot). We checked that the degeneracy holds for all non-Goldstone pion states. 
The lower left panel shows the infrared part of the directly calculated Dirac spectral density on the gauge
configurations and its Chebyshev expansion based approximation. The lower right panel shows the pion decay 
constant ${\rm F_\pi}$ fitted with the mixed action for fixed sea mass as a function of valence masses.
Continued future work is needed for definitive results of  BSM phenomenology. 
%
\section{The light $0^{++}$ scalar and the resonance spectrum}

The most important goals of our lattice Higgs project are
to establish the emergence of the light scalar state with $0^{++}$ quantum numbers and the resonance spectrum far
separated from the light composite scalar.

\subsection{The light scalar state}

The  ${\rm f_0}$ meson (in QCD terminology) has  ${\rm 0^{++}}$ quantum numbers and acts as the 
scalar state in the sextet model (${\rm \sigma}$ particle in QCD). 
Close to the conformal window, the ${\rm f_0}$ meson 
of the sextet model is not expected to be similar to its counterpart in QCD. If it turns out to be light, 
it can replace the elementary Higgs particle and pose as the Higgs impostor.  
Two types of different ${\rm 0^{++}}$ operators, the fermionic one and the gluonic one (${\rm 0^{++}}$ glueball), 
are expected to mix in the relevant correlation functions for mass determination. 
Such mixing was not included in the pilot study~\cite{Fodor:2014pqa} but becomes an important goal of our ongoing effort. 
We will report our new results without including these mixing effects.

A particular flavor-singlet correlator is needed to capture the ${\rm 0^{++}}$ scalar state with vacuum quantum numbers. It 
requires connected and disconnected diagrams of fermion loop propagators on ensemble gauge configurations.
The connected diagram corresponds to the non-singlet correlator  ${\rm C_{\rm non-singlet}(t)}$. 
The correlator of the disconnected diagram is  ${\rm D(t)}$ at time separation ${\rm t}$. 
The ${\rm f_0}$ correlator ${\rm C_{\rm singlet}(t)}$ is defined as ${\rm C_{\rm singlet}(t) \equiv C_{\rm non-singlet}(t) + D(t)}$. 
The transfer matrix has the spectral decomposition of the ${\rm C_{\rm singlet}(t)}$ correlator in terms of the 
sum of all energy levels ${\rm E_i(0^{++}), i=0,1,2,...}$ and their  parity partners ${\rm E_j(0^{-+}), j=0,1,2,...}$ 
but at large time separation ${\rm t}$ the lowest states ${\rm E_0(0^{++}) }$ and ${\rm E_0(0^{-+})}$ dominate. 
They correspond to ${\rm m_{f_0}}$ and ${\rm m_{\eta_{\rm sc}}}$. 
The relevant non-singlet 
staggered correlator can be fitted well with non-oscillating $a_0$ contribution and oscillating $\pi_{\rm sc}$ contribution, with the 
non-Goldstone pion $\pi_{\rm sc}$ discussed in Section 3. 
One of the most important new developments in our analysis is to use correlators which project out non-zero momentum 
states of the scalar. This projection eliminates the vacuum contribution in the disconnected part and improves 
the mass extraction procedure.

We estimate the connected and disconnected diagrams with stochastic source vectors of fermion propagators.
To evaluate the disconnected diagram, we  need to calculate closed loops of quark propagators. 
We introduce $Z_2$ noise sources on the lattice where each source is defined on individual time-slice $t_0$ for color $a$. 
The scheme  can be viewed as a ``dilution'' scheme which is fully diluted in time and color and even/odd diluted in space.
Results from the original pilot study~\cite{Fodor:2014pqa} 
on $32^3 \times 64$ lattice volumes at $\beta=3.20$ could only extend down to the lowest fermion mass 
at ${\rm m=0.003}$.
From our new analysis some representative examples  of ${\rm 0^{++}}$ effective mass fits  
are shown in Figure~\ref{fig:scalar} probing the light scalar closer to the chiral limit than before at fermion mass ${\rm m=0.0015}$.
The upper left panel is at $\beta=3.20$ with $48^3 \times 96$ lattice volume and the upper right panel is 
at $\beta=3.20$ with $56^3 \times 96$ lattice volume to check against finite volume dependence in this low 
fermion mass range. The two lower panels of the plot show results at $\beta=3.25$.

Although our original estimate ${\rm M_{0^{++}}/F \sim 1-3}$  for the chiral limit remains consistent with 
the ongoing new analysis, important further work is needed on the light ${\rm f_0}$ scalar with ${\rm 0^{++}}$ quantum numbers.
We want better control on the slowly changing topology of the RHMC algorithm and the related dependence of the 
extracted masses on the topological quantum numbers of the gauge configurations. We are also in the process 
of a closely related study of the $\eta'$ problem which is particularly interesting and important in the staggered fermion
formulation. 
Fermion mass deformations of the low-lying ${\rm f_0}$ state and the Goldstone pion are expected to be entangled
which requires extended ${\rm \chi PT}$ analysis. 
Our ongoing work will have to address these issues.

\subsection{The emerging resonance spectroscopy}

It is important to investigate the chiral limit of composite hadron states separated 
from the Goldstones and the light scalar by finite mass gaps. The baryon mass gap in the chiral limit, for example, 
provides further evidence
for ${\rm \chi SB}$ with preliminary results reported earlier~\cite{Santanu:2015}.  
Resonance masses of parity partners provide important additional information with split parity masses in the chiral limit.
This is particularly important for consistency with ${\rm \chi SB}$ and for a first estimate of the S parameter 
when probing the model via Electroweak precision tests~\cite{Peskin:1991sw}.

A remarkable resonance spectrum is emerging in our new analysis which is sketched in Figure~\ref{fig:spectroscopy} 
for illustration only. The scale is set by F in TeV units at both lattice spacings with caveats from discussions in 
Section 3 of the report.
Any conclusion about $\rm{\chi SB}$ or conformal behavior from eyeballed inspection of the data would be 
inappropriate and misleading. 
Although with more work needed for confirmation, the sextet model appears to be close to the conformal window and 
due to $\chi{\rm SB}$ exhibits
the right Goldstone spectrum for the minimal realization of the composite Higgs mechanism with a light scalar 
separated from the associated resonance spectrum in the 2-3 TeV region.
Chiral symmetry breaking and a very small beta function are
not sufficient to guarantee a light dilaton-like state as the natural interpretation for the emergence of the light scalar.
Consistent with our observations, a light Higgs-like scalar is still expected to emerge near the conformal window as a composite state with $0^{++}$ quantum numbers, but not necessarily with a dilaton interpretation.
This scalar state has to be light but is not required to match exactly the observed ${\rm 125~GeV}$
mass. The light scalar
from composite strong dynamics gets lighter
from electroweak loop corrections,  dominated by the large negative mass shift from 
the top quark loop~\cite{Foadi:2012bb,Cacciapaglia:2014uja,DiChiara:2014uwa}.

\section{The scale dependent renormalized coupling and beta function}

An important and independent consistency condition of the model would be provided
by matching the scale dependent renormalized coupling of the perturbative regime to the scale dependent 
coupling of the non-perturbative phase associated with  $\chi{\rm SB}$. 
We proposed a gauge coupling earlier
$g(\mu  = 1/L)$, running with the scale set by the finite volume~\cite{Fodor:2012td}
and defined on the gradient flow of the gauge field~\cite{Luscher:2010iy}.  
Since the gradient flow at flow time $t$ probes the gauge field at the scale $\sqrt{8t}$, the running coupling can be defined as a
function of $L$ in finite volume $V=L^4$ while holding $c=(8t)^{1/2}/L$  fixed with
%
${\rm \alpha_{c}(L)=4\pi\langle t^2E(t)\rangle /[3(1+\delta (c)]}$ 
with details of the definition given in ~\cite{Fodor:2012td}.
%
%
%
This volume dependent coupling is 
particularly suitable to study the perturbative
regime and departures from it.
The measured renormalized couplings are very accurate and  the scheme defines a one-parameter family when $c$
is varied and adjusted for different goals~\cite{Fodor:2014cpa,Fodor:2014cxa}.

For illustration, published results are presented in Figure~\ref{fig:beta12} to show the efficiency of the method. 
For comparison, red color marks the step $\beta$-function of four massless fermion flavors in the fundamental representation
from an earlier publication~\cite{Fodor:2014cpa}. The new published result for the step $\beta$-function of eight massless fermion flavors in the fundamental representation is marked with green color in the plot~\cite{Fodor:2015baa}. 
The step $\beta$-function of the sextet model of this report is marked with blue color in the plot from a 
recent publication~\cite{Fodor:2015zna}. This result ruled out an earlier claim of an infrared fixed point (IRFP) in the 
${\rm g^2 = 6}$ range of the renormalized coupling which would have implied a conformal 
phase for the sextet model~\cite{Hasenfratz:2015ssa}.

It is important to note that an IRFP  was reported in the ${\rm g^2=6.0-6.4}$ 
range of the renormalized coupling with twelve massless fermions in the fundamental representation~\cite{Cheng:2014jba}.
This IRFP is based on the study of the scale-dependent running coupling from the gradient 
flow with the choice $c=0.2$ and has been generally accepted 
as the most convincing evidence that the model with twelve massless flavors in the fundamental representation 
is inside the conformal window. Recent work using larger $c$ values in the gradient flow scheme 
does not find an IRFP in the reported range~\cite{Lin:2015zpa}. Further work would be required to resolve the tension 
between these two findings.

To establish the volume dependent running coupling without an IRFP in the range of renormalized couplings
which can be reached within limitations of the method should be only one important part of the analysis. 
Equally important is to find independent and direct evidence for ${\rm \chi SB}$ to rule out the conformal phase.
In this report we presented a preponderance of evidence for ${\rm \chi SB}$ in the sextet model. 
For added consistency,
in the non-perturbative phase with ${\rm \chi SB}$ we are interested in a scale-dependent 
and volume independent renormalized coupling.
At fixed lattice size, bare coupling, and fermion mass m
we determine the appropriate flow time ${\rm  t(g^2, m)}$ to match any targeted 
flow-dependent renormalized coupling ${\rm g^2}$ calculated from ${\rm \langle t^2E(t)\rangle}$.
Assuming that the footprint of the operator  on the gradient flow is sufficiently small 
compared to the Compton wavelength of the pion for p-regime
analysis, the dependence of  ${\rm  t(g^2, m)}$ on ${\rm m}$ can be replaced by ${\rm  t(g^2, M^2_\pi)}$ in 
${\rm \chi PT}$ of pion dynamics with
linear dependence of  ${\rm  t(g^2, M^2_\pi)}$ on ${\rm M^2_\pi}$ in leading order~\cite{Bar:2013ora}.
Any residual finite volume dependence can be corrected in ${\rm \chi PT}$.
A scale dependent and volume independent step 
beta function can be determined from this procedure in the phase with ${\rm \chi SB}$. 
The analysis is far from complete in the sextet model. Plans for similar calculations in the fundamental rep 
with twelve flavors, motivated by our newest results, would be interesting to consider.

\section{ Computational framework}

We developed promising and well-tested new methods  to overcome some limitations of more conventional 
lattice BSM methods. They include the analysis of Goldstone dynamics with mixed actions, the chiral condensate, and spectroscopy 
in crossover from the p-regime to the $\epsilon$-regime, close to the chiral limit of the theory.
As a candidate theory of the composite Higgs particle
the new approach requires new runs and a mixed valence action with excellent chiral properties of valence fermions. 
The required fermion mass range of the gauge configurations is below the range of our existing run set. 
Runs in the range from  ${\rm m=0.0010}$ to ${\rm m=0.0003}$ operate in the crossover region from
the p-regime to the $\epsilon$-regime and runs with valence fermions in the 
${\rm m=0.0003-0.0001}$ range reach further down in the fermion mass range of the  RMT analysis.
Three gauge couplings are needed to control the continuum limit. 

Our physics analysis includes spectroscopy, the Dirac spectra, and
the running coupling in capacity computing mode with new configuration generation plans.
The disconnected diagrams of the low mass ${\rm 0^{++}}$ state are extremely demanding on gpu
resources. The conventional spectroscopy including analysis with a series of 
valence fermion masses in mixed actions, 
Dirac spectra including RMT, and configuration generation for the volume dependent running coupling also
require large gpu resources for every given run parameter set in large volumes. 

We use the tree-level Symanzik-improved gauge action for all simulations reported here.
The conventional $\rm{\beta=6/g^2}$ lattice gauge coupling is defined as the overall
factor in front of the well-known terms of the Symanzik lattice action. 
Its values are in the $\beta=3.20-3.30$ range of the proposed ensemble generation.
The link variables in the staggered fermion matrix are exponentially smeared with  two
stout steps~\cite{Morningstar:2003gk}; the precise definition of the staggered stout action was given in~\cite{Aoki:2005vt}.  
The RHMC algorithm is deployed in all runs. The fermion flavor doublet requires rooting in the algorithm.
For molecular dynamics time evolution we apply multiple time scales~\cite{Urbach:2005ji} and the
Omelyan integrator~\cite{Takaishi:2005tz}.
Our error analysis of  hadron masses uses correlated fitting with double jackknife procedure on the covariance matrices.
The time histories of the fermion condensate, the plaquette, the topological charge,
and correlators are used to monitor autocorrelation times in the simulations.

The composite Higgs theory of this report requires a code structure which is distinct from lattice QCD in one important way.
The fermions in QCD are in the fundamental representation of the SU(3) gauge group and represented in the code by a 
complex vector ${\rm q_i}$, with  color index ${\rm i=1,2, 3}$. In the sextet model the fermions are represented as a two-index 
symmetric tensor $T_{ij}$ which is constructed as the symmetric part of the tensor ${\rm q_iq_j}$. 

The staggered Dirac operator, represented in matrix form and built from the SU(3) gauge link variables, 
has to be applied to the fermions 
field during the CG iteration steps consuming the dominant part of the ensemble generating evolution code.
We developed two versions of the code. In the vector code, using the 6-dimensional vector representation of the fermions, 
the SU(3) gauge link matrices $U_{ij}$ are brought in from main memory and from the product $U_{ij}\cdot U_{kl}$ on each link a new 
6x6 matrix $V_{mn},~m,n=1,2,...6$ is constructed which acts on the $Q_n$ vectors as a $V_{mn}\cdot Q_n$ matrix-vector product.
It is possible, however, to keep the Dirac matrix-vector algebra in the original two-index symmetric representation which entirely eliminates
the overhead required to reconstruct the 6x6 matrix $V_{mn}$  and directly use the product of SU(3) link matrices 
on each link in tensor representation.
The load on the memory bandwidth remains unchanged and on the BG/Q platform 
with excellent balance of the network bandwidth and 
memory bandwidth this leads to a factor of two reduction in flops and to a corresponding speedup factor in performance.

The code, with high utilization of the fast BG/Q network and with register level optimization of the local code on the nodes,
was thoroughly tested for weak scaling and strong scaling across a  large number of nodes on the BG/Q platform. 
The efficiency of the code in CG performance is 29 percent of the peak speed of the BG/Q platform 
on ${\rm 48^3\times 96}$ lattices on one rack with little loss of efficiency on half-rack. 
The numbers illustrate our efficient BG/Q code in lattice gauge theory applications of staggered fermions.

\section{Acknowledgement}


We acknowledge support by the DOE under grant DE-SC0009919,
by the NSF under grants 0970137 and 1318220, by the DOE ALCC award for the BG/Q Mira platform
of Argonne National Laboratory, by OTKA under the grant OTKA-NF-104034, and by the Deutsche
Forschungsgemeinschaft grant SFB-TR 55. Computational resources were provided by the Argonne Leadership Computing Facility under an ALCC award, by USQCD at Fermilab, by the University of Wuppertal, by The Juelich Supercomputing Center on Juqueen
and by the Institute for Theoretical Physics, Eotvos University. 
We are grateful to Szabolcs Borsanyi for his code development for the BG/Q platform. We are also 
grateful to Sandor Katz and Kalman Szabo for their code development for the CUDA platform \cite{Egri:2006zm}. KH wishes to thank the Institute for Theoretical Physics and the Albert Einstein Center for Fundamental Physics at the University of Bern for their support.


\clearpage

{\bf\Large Figures}
\vskip 0.1in
\setcounter{figure}{0}
%
%


%
\begin{figure}[ht!]
	\begin{center}
		\begin{tabular}{c}
			\includegraphics[width=6.5in]{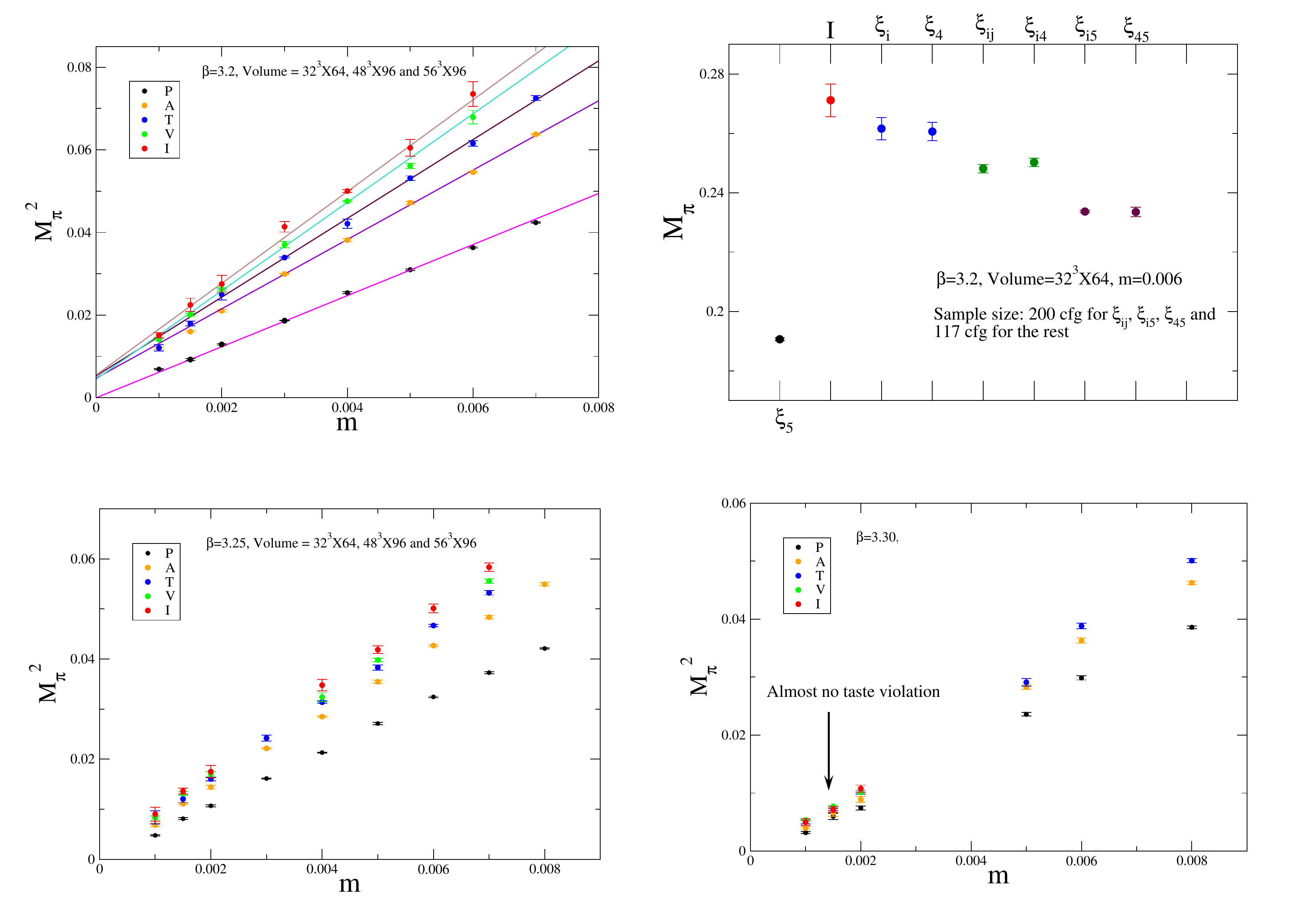}
		\end{tabular}
	\end{center}
	\caption{\footnotesize  
		To illustrate cutoff dependent taste breaking effects, spectra of mass-deformed non-Goldstone pion states are shown
		from our newest data with the definition of the relevant correlators and
		quantum numbers given in~\cite{Fodor:2011tu,Fodor:2012ty}.
		In the fermion mass range of our data set the taste breaking pattern 
		is different from QCD where the residual ${\rm \Delta}$ mass shifts of the non-Goldstone pions are equispaced
		in the chiral limit with approximately degenerate SO(4) taste multiplets and with  parallel slopes for finite fermion mass deformations 
		of Goldstone and non-Goldstone pion states~\cite{Lee:1999zxa}.
		For example, as part of the equispaced split of degenerate SO(4) multiplets, the observed approximate split
		${\rm \Delta_{ij} \sim  2\Delta_{sc}}$ of two multiplets in QCD appears to have collapsed in the sextet model.
		The other distinct difference from QCD is the non-parallel slopes which fan out  
		in Goldstone and non-Goldstone  mass deformations of the pion spectrum as shown.
		While the ${\rm \Delta}$ additive mass shifts are LO taste breaking effects 
		in the chiral Lagrangian~\cite{Lee:1999zxa,Aubin:2003mg}, the taste breaking slope corrections ${\rm \delta}$
		can plausibly be identified with NLO analytic terms in the chiral analysis~\cite{Sharpe:2004is}. The corrected mass 
		relation is ${\rm M^2_{NLO} = M^2_{LO}(1+\delta)}$
		where ${\rm \delta}$ depends on the taste quantum number of the pion state. 
		Several relations constrain the $ \delta$ taste breaking corrections~\cite{Sharpe:2004is}.}
	
\label{fig:goldstone}	
\end{figure}

\vskip 0.25in

\begin{figure}[htb!]
	\begin{center}
		\begin{tabular}{cc}
			\includegraphics[height=6cm]{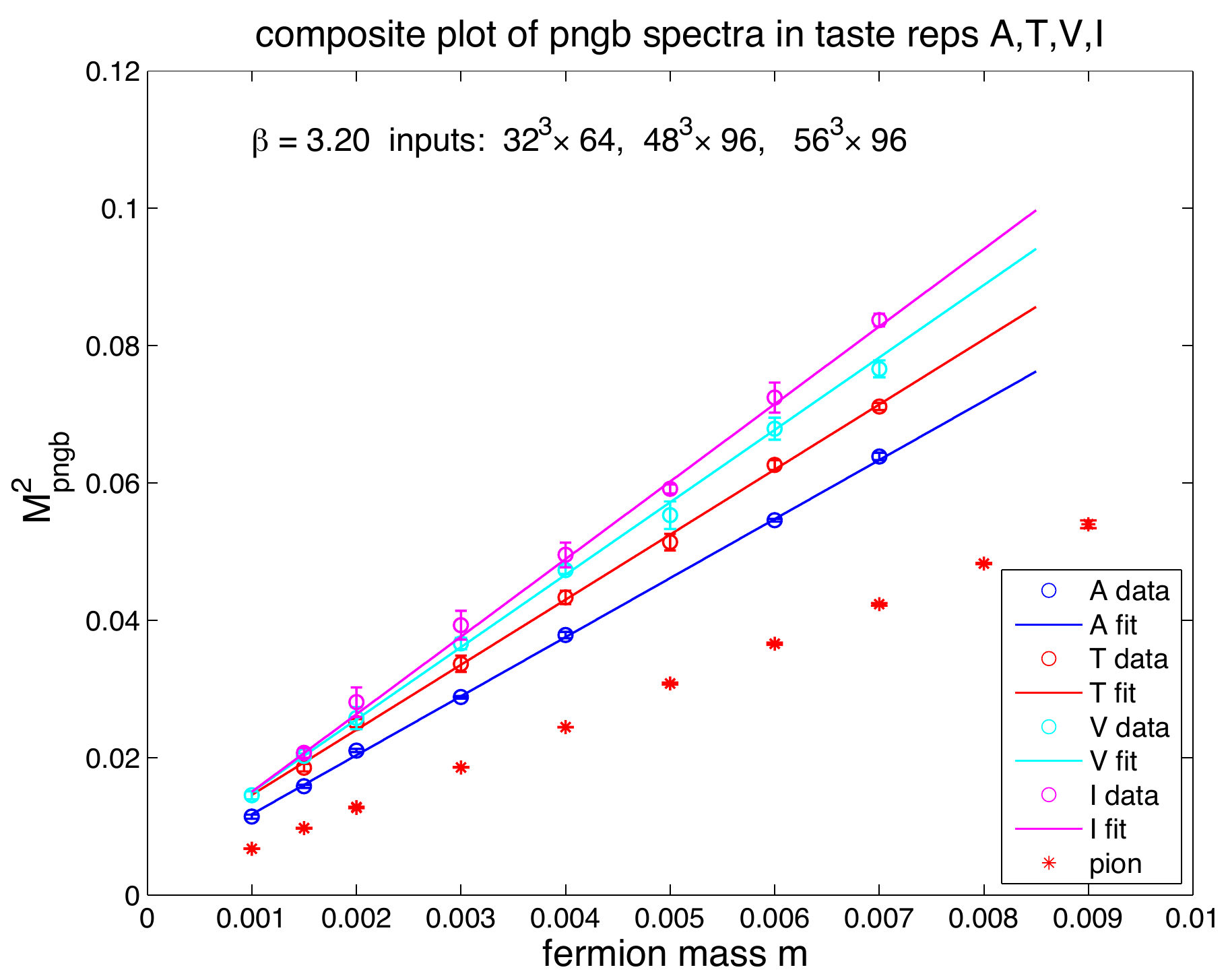}&
			\includegraphics[height=6cm]{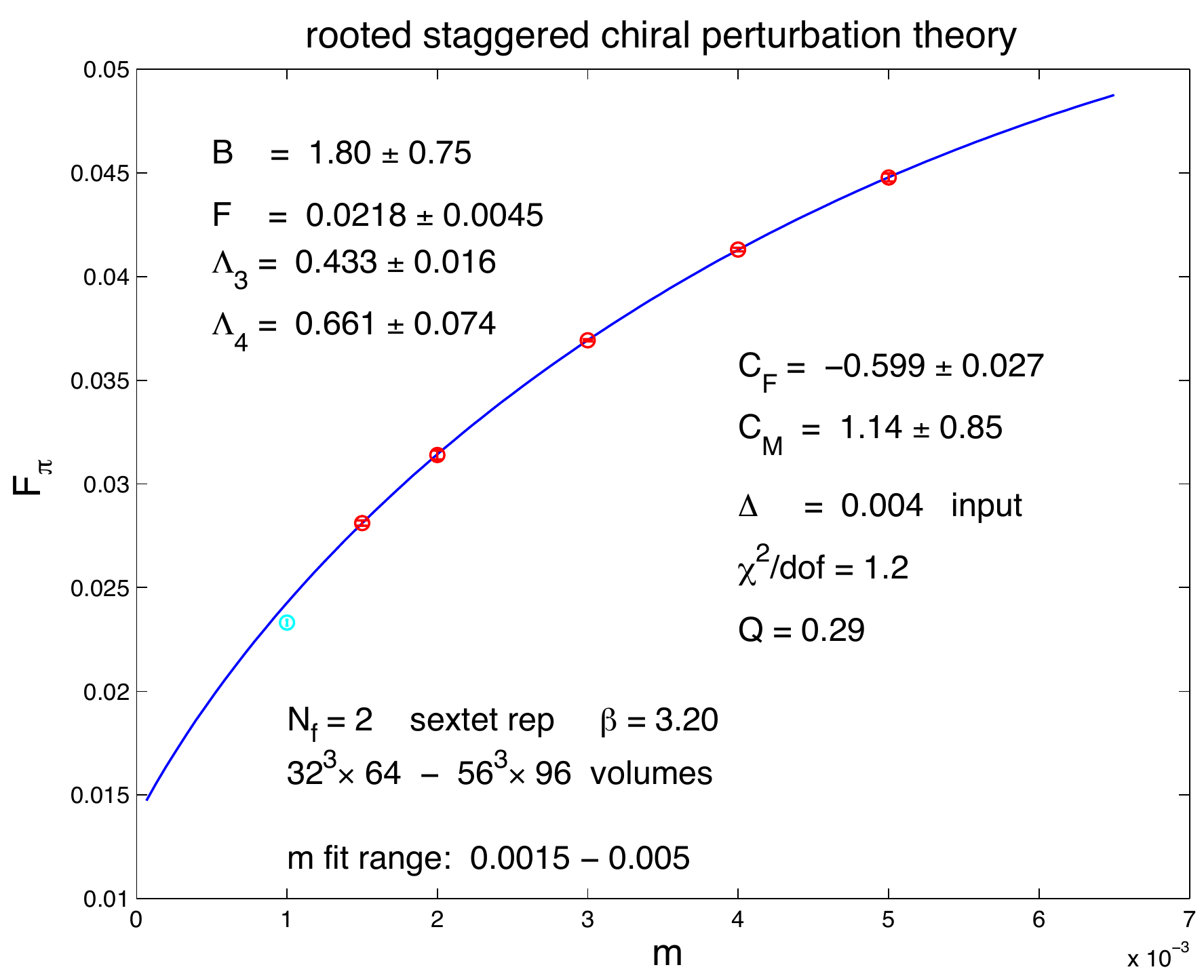}\\
			\includegraphics[height=6cm]{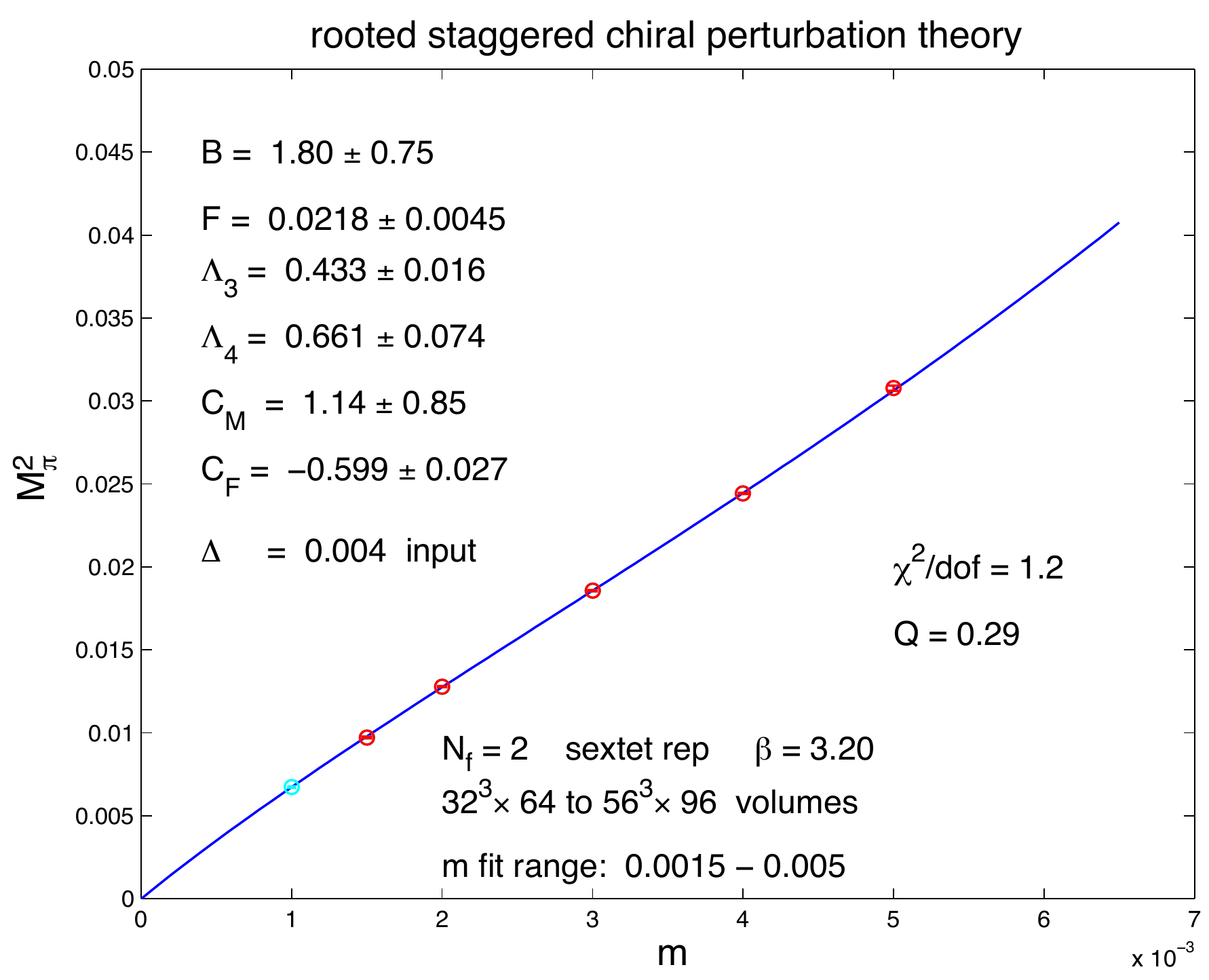}&
			\includegraphics[height=6cm]{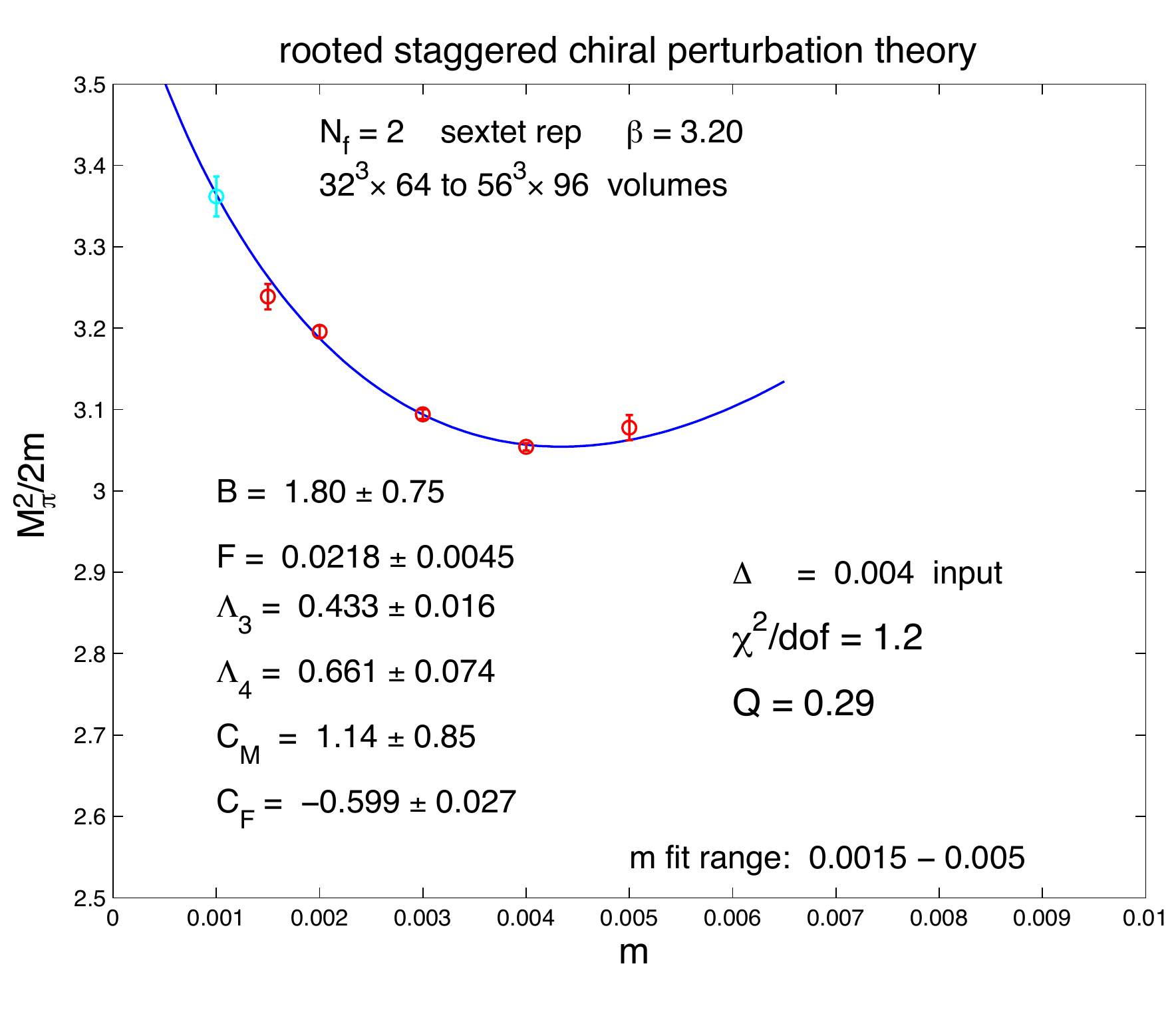}
		\end{tabular}
	\end{center}
	\caption{\footnotesize  Preliminary results from rooted ${\rm \chi PT}$ are shown
		from fits at gauge coupling ${\rm \beta = 3.20}$ 
		which corresponds to our coarser lattice of the two extended sets of gauge ensembles. 
		The upper left panel shows the linear fits to the quadratic masses of the non-Goldstone pions to determine their 
		mass shifts and slopes as input. The upper right panel shows the 
		rooted ${\rm \chi PT}$ fit to ${\rm F_\pi}$ as a function of fermion mass
		deformations away from the chiral limit. The two lower panels  show  
		rooted ${\rm \chi PT}$ fits to ${\rm M_\pi}$ as a function of fermion mass
		deformations away from the chiral limit. 
		We have similar analysis for ${\rm M^2_{\pi}}$ and ${\rm F_{\pi}}$ at ${\rm \beta=3.25}$. }
	\label{fig:chipt}
\end{figure}
\begin{figure}[ht!]
	\begin{center}
		\begin{tabular}{c}
			\includegraphics[width=6.5in]{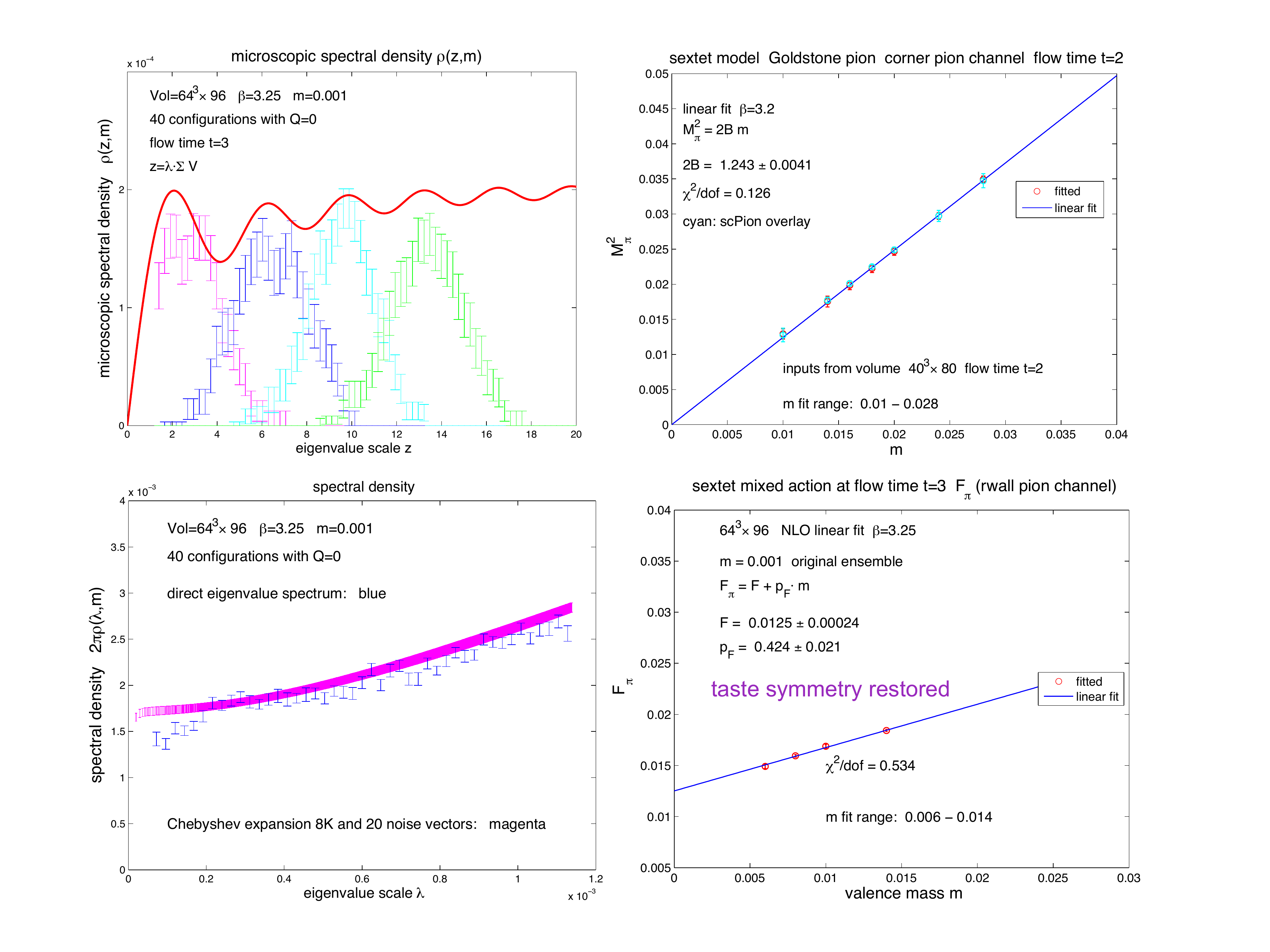}
		\end{tabular}
	\end{center}
	\caption{\footnotesize  The newest test results are shown in the plot.
		The valence action is defined with a large number of small stout steps which corresponds 
		to gradient flow time ${\rm t=3}$ in cutoff units at each gauge coupling.
		We checked the eigenvalues of the Dirac operators which order into nearly degenerate quartets 
		with the smeared gauge links of the gradient flow. The degenerate eigenvalues follow the index theorem count matching the 
		topology of each gauge configuration as measured from the topological charge operator on the gradient flow. 		
		The upper left panel shows the RMT distribution of four degenerate quartets showing that
			the $\epsilon$-regime is reached with the scaling variable ${\rm \lambda\Sigma_{mixed} V\sim 10}$ 
			where the fermion mass is replaced by the scale 
			of the gradient flow defined valence Dirac spectrum (${\rm m\rightarrow\lambda)}$.
			The fermion condensate  ${\rm \Sigma_{mixed}}$, not RG invariant itself, is 
			consistently determined from the gradient flow defined valence Dirac operator.
		The upper right panel illustrates the perfect degeneracy of the Goldstone pion with one selected non-Goldstone pion 
		(scPion in the plot). We checked that the degeneracy holds for all non-Goldstone pion states. 
		The lower left panel shows the infrared part of the directly calculated Dirac spectral density on the gauge
		configurations and its Chebyshev expansion based approximation. The lower right panel shows the pion decay 
		constant ${\rm F_\pi}$ fitted with the mixed action for fixed sea mass as a function of valence masses.	}
	\label{fig:mixed}
\end{figure}

\vskip 0.25in

\begin{figure}[htb!]
	\begin{center}
		\begin{tabular}{cc}
			\includegraphics[height=6cm]{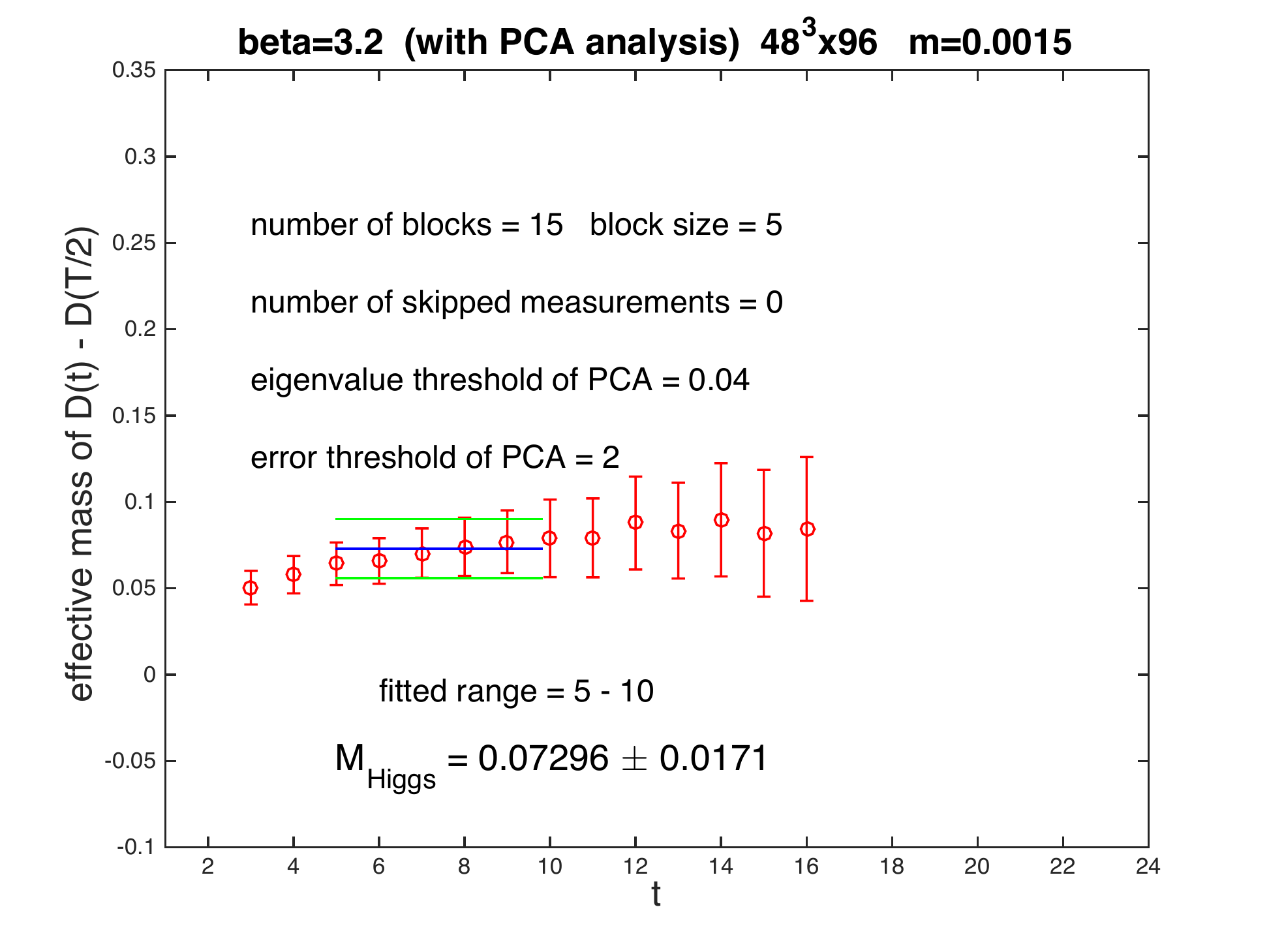}&
			\includegraphics[height=6cm]{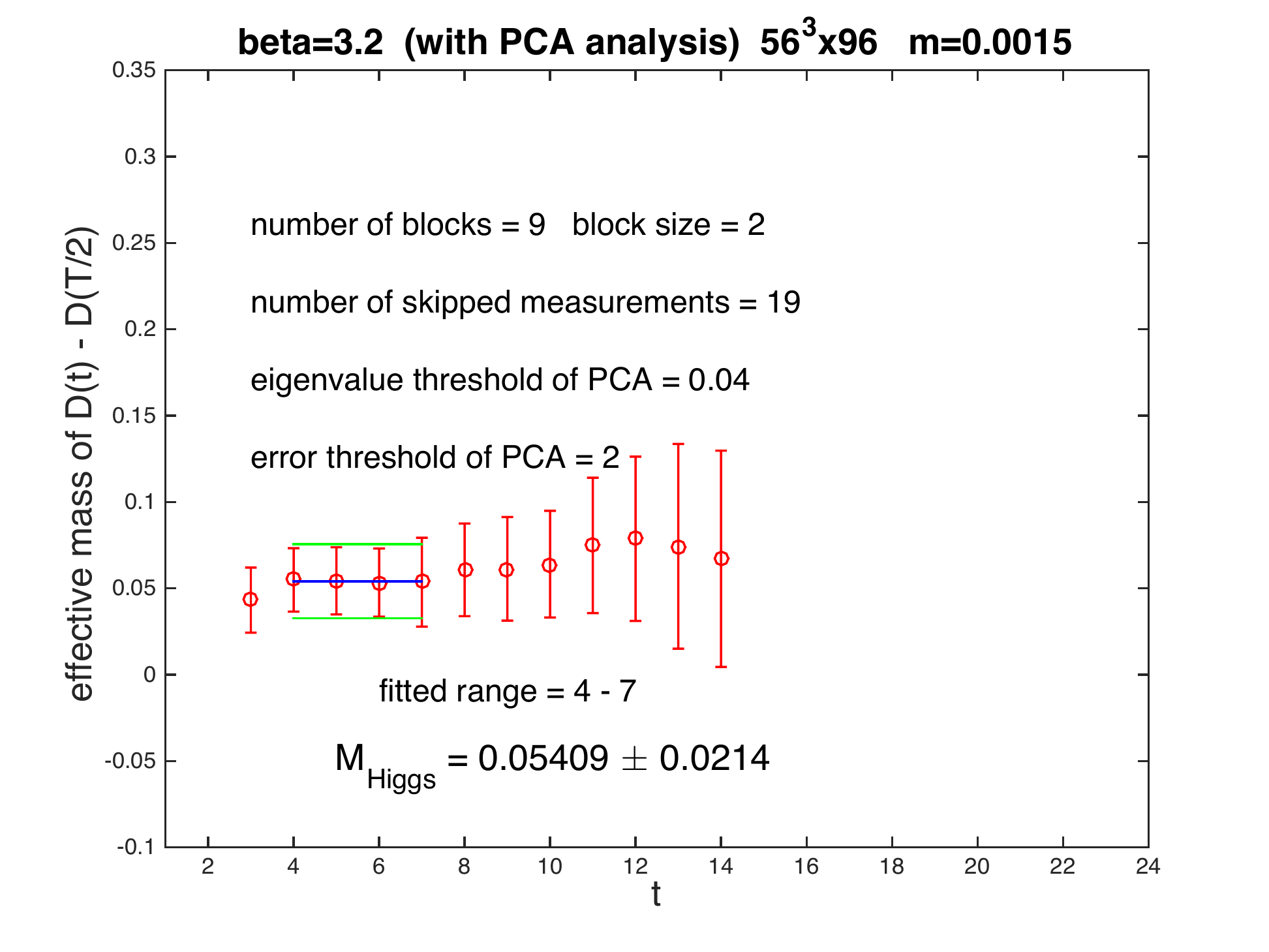}\\
			\includegraphics[height=6cm]{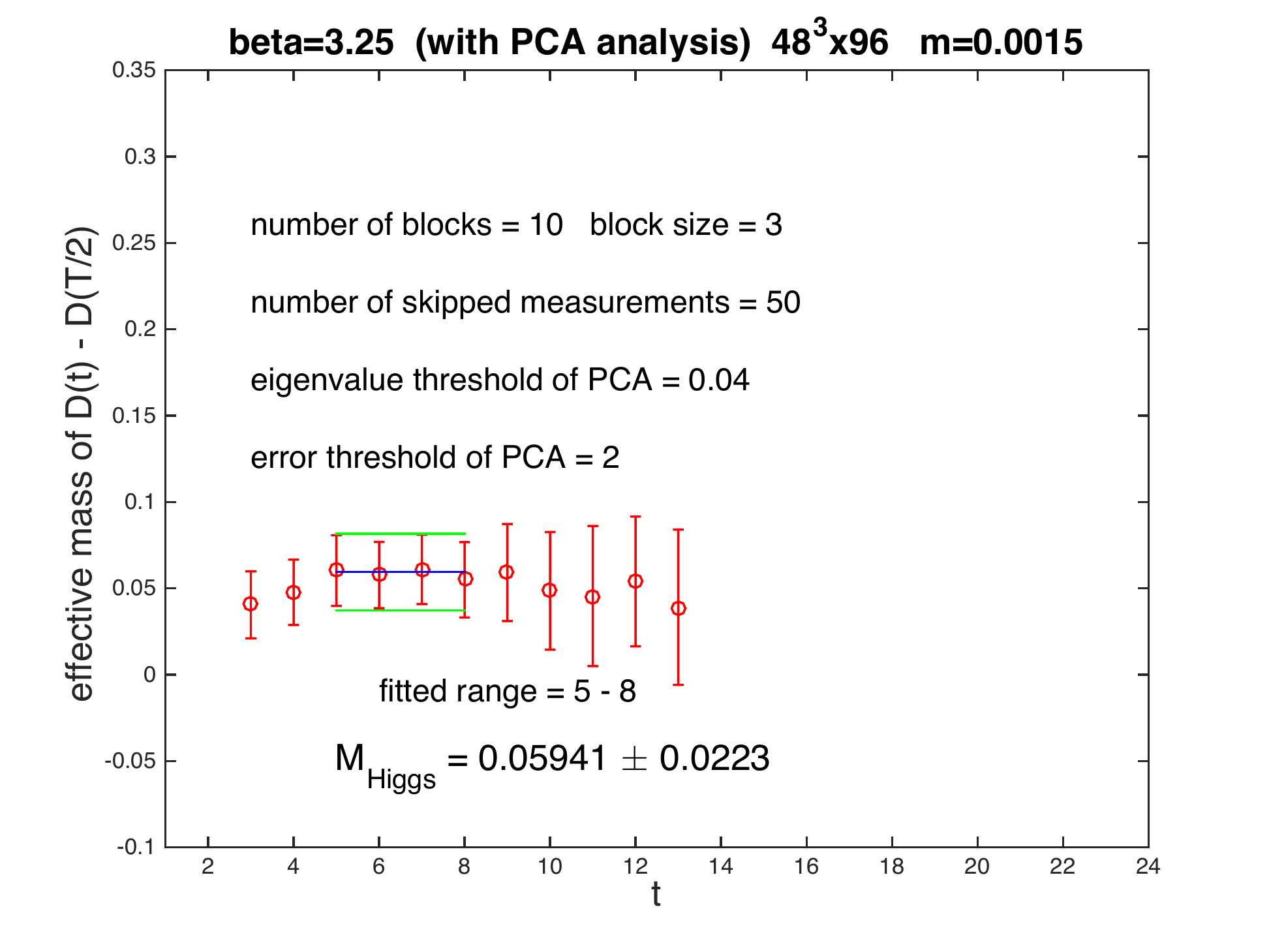}&
			\includegraphics[height=6cm]{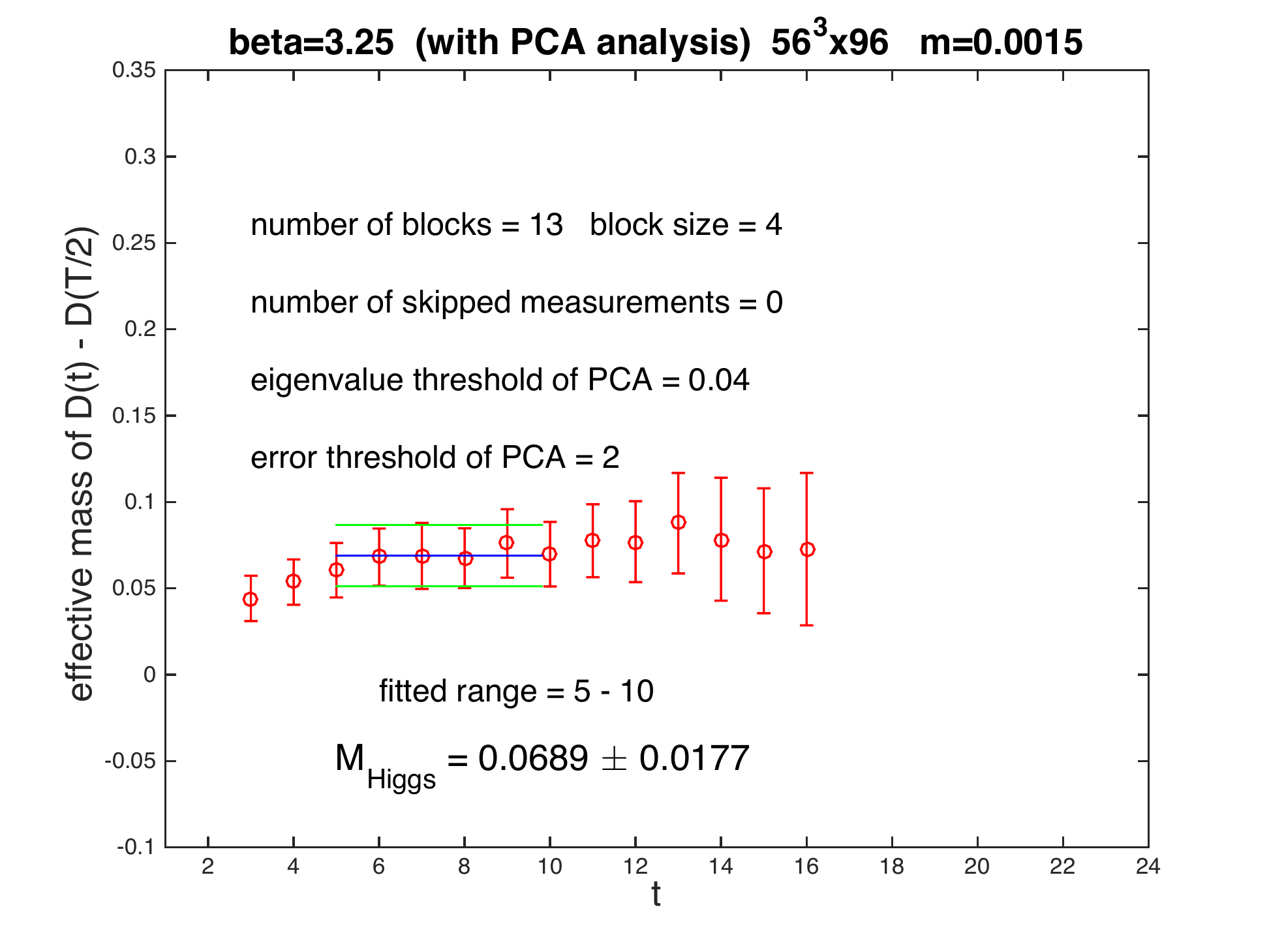}
		\end{tabular}
	\end{center}
	\caption{\footnotesize  Representative fits of the low mass scalar from two ensembles using double Jackknife  
			procedure on the covariance matrix with Principal Component Analysis (PCA). }
	\label{fig:scalar}
\end{figure}
\begin{figure}[ht!]
	\begin{center}
		\begin{tabular}{c}
			\includegraphics[width=6.5in]{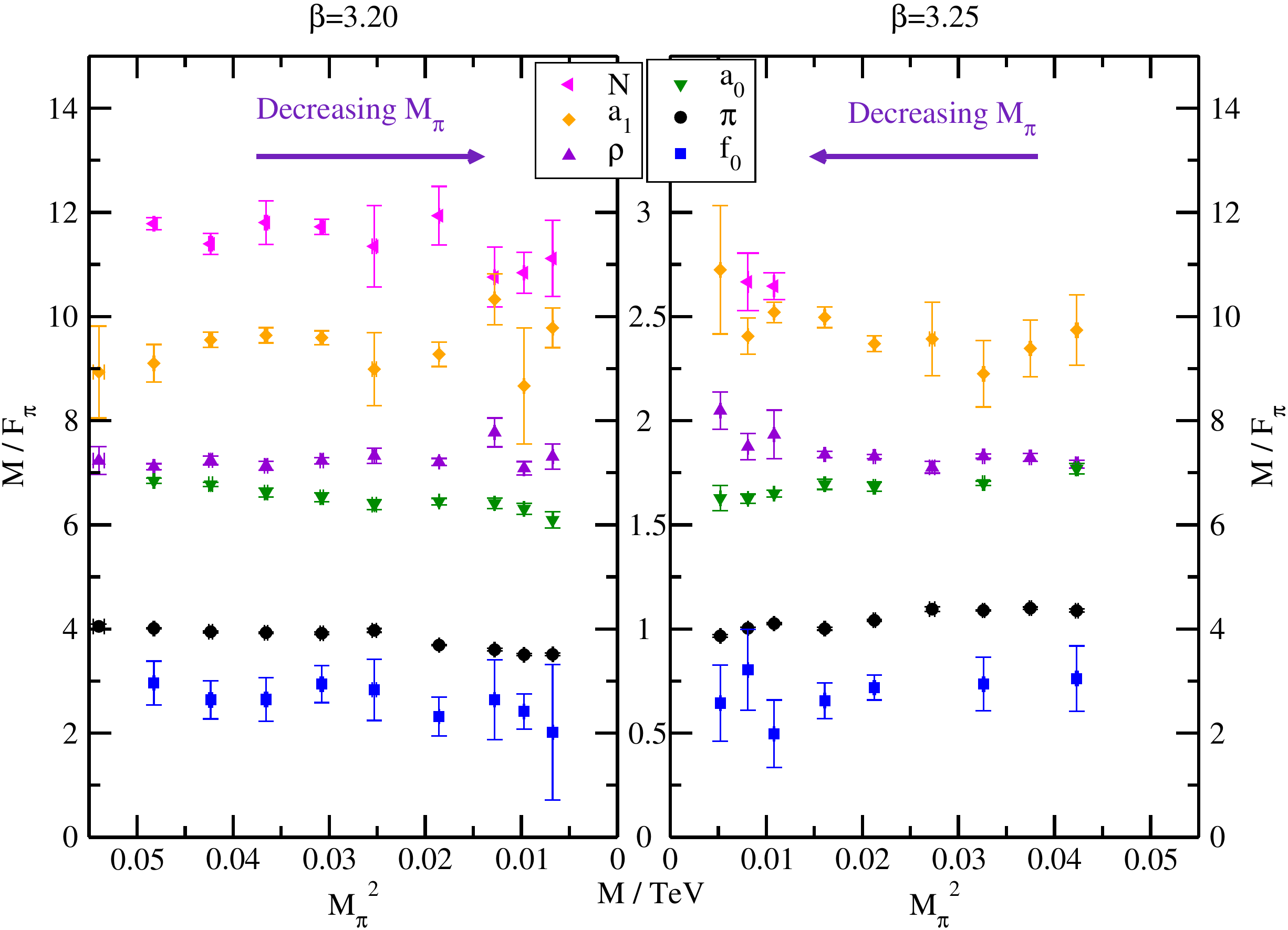}
		\end{tabular}
	\end{center}
	\caption{\footnotesize  New resonance spectroscopy results are shown in the plot for illustration only. 
		The scale is set by F=250 GeV at both lattice spacings with caveats from discussions in 
		Section 3 of the report.
		Any conclusion about $\rm{ \chi SB}$ or conformal behavior from
		eyeballed inspection of the data would be inappropriate and misleading. }
	\label{fig:spectroscopy}
\end{figure}
\begin{figure}[ht!]
	\begin{center}
		\begin{tabular}{c}
			\includegraphics[width=6.5in]{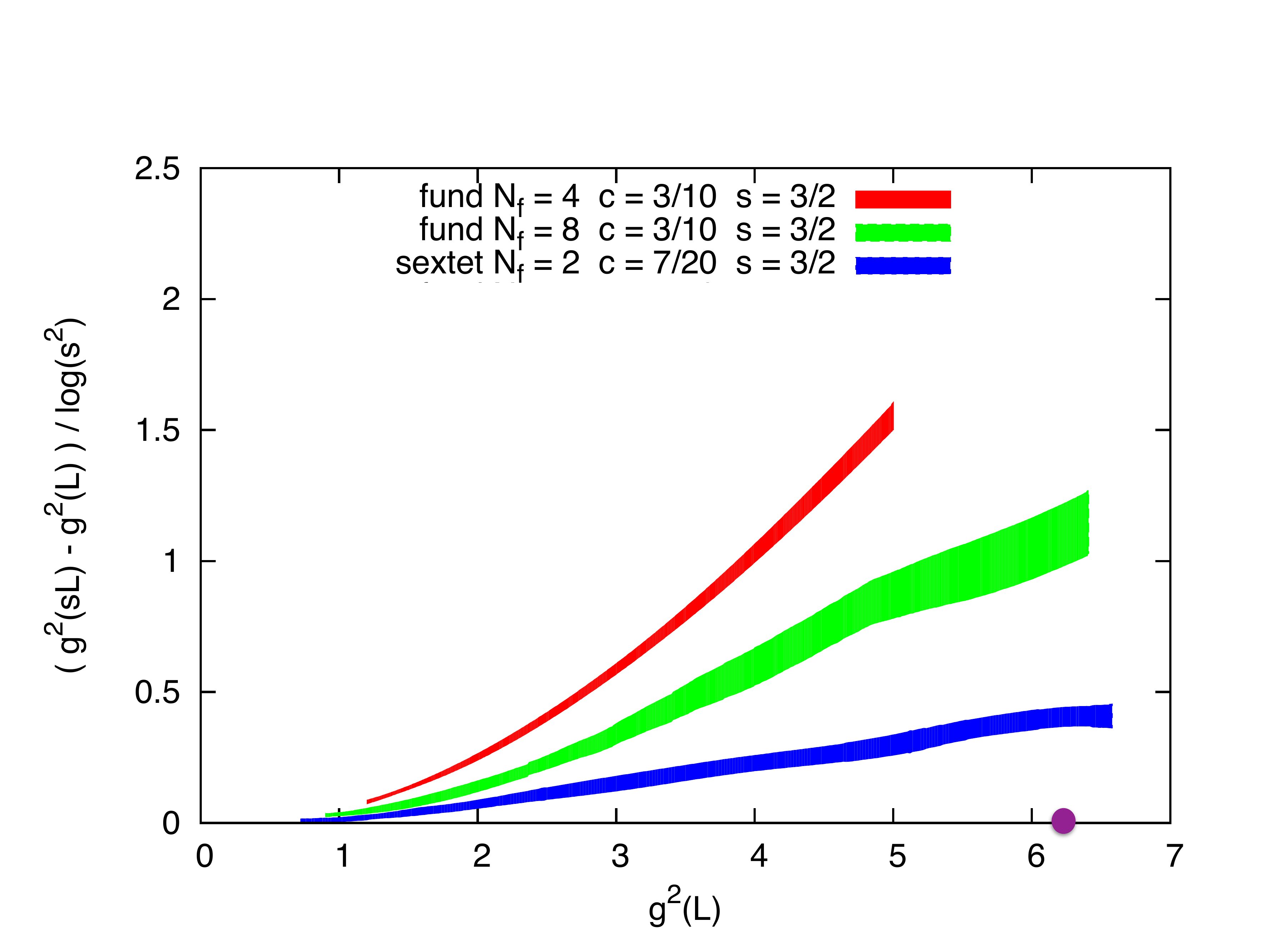}
		\end{tabular}
	\end{center}
	\caption{\footnotesize  Comparative results are shown in the plot to illustrate the efficiency of the gradient flow based running coupling and $\beta$-function.
		For comparison, red color marks the step $\beta$-function of four massless fermion flavors in the fundamental representation
		from an earlier publication~\cite{Fodor:2014cpa}. The new published result for the step $\beta$-function of eight massless fermion flavors in the fundamental representation is marked with green color in the plot~\cite{Fodor:2015baa}. 
		The step $\beta$-function of the sextet model of this proposal is marked with blue color in the plot from a 
		recent publication~\cite{Fodor:2015zna}. This result ruled out an earlier claim of an infrared fixed point (IRFP) in the 
		${\rm g^2 = 6}$ range of the renormalized coupling which would have implied conformal phase for the sextet model.
       The purple point in the plot marks the IRFP  reported in the renormalized coupling with twelve 
	massless fermions in the fundamental representation~\cite{Cheng:2014jba}.		}
	\label{fig:beta12}
\end{figure}

\end{document}